\documentclass[twocolumn,prl,aps,showpacs]{revtex4}

\usepackage{graphicx}
\usepackage{dcolumn}
\usepackage{bm}

\def\be{\begin{equation}}
\def\ee{\end{equation}}
\def\bea{\begin{eqnarray}}
\def\eea{\end{eqnarray}}

\begin{document}

\title{$Q^2$ Evolution of Generalized Baldin Sum Rule for the Proton}

\author{
Y.~Liang,$^{1}$
M.E.~Christy,$^{2}$ R.~Ent,$^{3}$ C.E.~Keppel$^{2,3}$}
 \affiliation{
$^{1}$ Ohio University.
$^{2}$ Hampton University.
$^{3}$ Thomas Jefferson National Accelerator Facility.}
\newpage
\date{\today}

\begin{abstract}
The generalized Baldin sum rule for virtual photon scattering, the unpolarized analogy of the generalized Gerasimov-Drell-Hearn integral, provides an important way to investigate the transition between perturbative QCD and hadronic descriptions of nucleon structure. This sum rule requires integration of the nucleon structure function $F_1$, which until recently had not been measured at low $Q^2$ and large $x$, i.e. in the nucleon resonance region. This work uses new data from inclusive electron-proton scattering in the resonance region obtained at Jefferson Lab, in combination with SLAC deep inelastic scattering data, to present first precision measurements of the generalized Baldin integral for the proton in the $Q^2$ range of 0.3 to 4.0~GeV$^2$.
\end{abstract}


\maketitle

Polarizabilities are the fundamental quantities that characterize the response of a composite system to static or slowly-varying external electromagnetic fields. The Baldin sum rule connects the sum of the electric and magnetic polarizabilities of the nucleon ($\alpha+\beta$) to the integral of the $\nu^2$-weighted nucleon unpolarized photoabsorption cross section \cite{baldin,lapidus}, 
\begin{equation}
  \label{eq:baldin}
 {\alpha+\beta} ={{1\over {4\pi^2}}{\int_{\nu_0}^{\infty}{{\sigma_{1\over2}+\sigma_{3\over2}}\over {\nu^2}}\, d\nu}}.   
\end{equation}
Here, $\sigma_{1\over2}$ and $\sigma_{3\over2}$ are the photoabsorption cross sections of 1/2 and 3/2 helicity states, respectively; $\nu$ is the energy carried by the photon; and $\nu_0$ is the pion photoproduction threshold energy. The  polarizabilities $\alpha$ and $\beta$ are defined in the low energy expansion of the Compton scattering amplitudes. In particular, $\alpha+\beta$ represents the helicity non-flip part of the electromagnetic polarizability. The Baldin sum rule establishes a relation between the static nucleon properties (electric and magnetic polarizabilities) and the dynamic nucleon excitation spectrum, such that these polarizabilities can be extracted from precision measurements of the photoabsorption cross sections in real Compton scattering. For the proton, recent measurements give $(\alpha+\beta)_p$ $=13.69\pm0.14$ \cite{babusci}.

The Baldin sum rule is the unpolarized analog of the Gerasimov-Drell-Hearn (GDH) sum rule \cite{gerasimov,drell} and, in analogy to the generalized GDH sum rule \cite{anselmino}, D. Drechsel, B. Pasquini, and M. Vanderhaeghen have used dispersion relation formalism \cite{kronig,low} to extend the Baldin sum rule to virtual Compton scattering ($Q^2>0$, where $Q^2$ is the square of four-momentum transfer) \cite{drechsel}. This provides a tool to extract generalized polarizabilities by means of radiative electron scattering. These generalized polarizabilities are functions of the $Q^2$ of the incident photon and describe, in some sense, the spatial distribution of the polarizabilities. After a proof of principle at SLAC \cite{brand}, the first unpolarized virtual Compton scattering observables have been obtained from MAMI at $Q^2 = 0.33$~GeV$^2$ \cite{roche}, and recently at Jefferson Lab at higher $Q^2$ ($1< Q^2 <2$~GeV$^2$) \cite{jaminion,laveissiere}.

The virtual Compton scattering process includes the absorption of the virtual photon, which is related to inclusive electron-nucleon scattering. At finite $Q^2$, the generalized sum rule gives       
\begin{eqnarray}
  \label{eq:exbaldin}
 {\alpha(Q^2)+\beta(Q^2)} &=&{{1\over {4\pi^2}}{\int_{\nu_0}^{\infty}{K\over \nu}{{\sigma_{1\over2}+\sigma_{3\over2}}\over {\nu^2}}\, d\nu}} \nonumber\\
&=&{{{e^2M}\over {\pi{Q^4}}}\int_{0}^{x_0}{2xF_1(x,Q^2)}\, dx},   
\end{eqnarray}
where the integral on the right hand side is the second Cornwall-Norton moment \cite{cornwall} of the nucleon structure function $F_1$, barring the elastic contribution. Here, $M$ is the nucleon mass. According to Hand's definition \cite{hand}, $K=(W^2-M^2)/2M$ is the equivalent real photon energy needed to excite the nucleon to mass $W$. The Bjorken scaling variable is $x=Q^2/2M\nu$, and $x_0$ corresponds to pion threshold. 

In the limit of large $Q^2$, the coupling constant of Quantum ChromoDynamics (QCD) is very small, and perturbative QCD provides an excellent interpretation of the deep inelastic scattering (DIS) process of electron-proton scattering. At large $Q^2$, the Callan-Gross relation ($2xF_1 = F_2$) \cite{callan} is valid, and the second moment of the structure function $F_1$ is approximately equal to the first moment of the structure function $F_2$ which is nearly a constant at large $Q^2$ \cite{rujula, rujula1}. Therefore, the integrant in Eq. \ref{eq:exbaldin} will be constant, and the generalized sum rule ($\alpha+\beta) \sim 1/Q^4$, and will go to $0$ as $Q^2$ goes to $\infty$. At the larger distance scales probed at low $Q^2$, the coupling constant of QCD is large, and the scattering process is better described in terms of hadronic degrees of freedom using Chiral Perturbation Theory. The generalized sum rule, then, tends to the Baldin sum rule of real Compton scattering at $Q^2=0$. 

Between these two regions, a rigorously descriptive theory is lacking at present. Here, the sum rule is dominated by the resonance region structure function. Measuring the generalized Baldin sum rule in this transition region ($Q^2$ up to a few GeV) provides an unique window to understand the transition from DIS incoherent processes to the resonance-dominated coherent processes. In practice, we measure the second moment of $F_1$ in the inclusive electroproduction process, and then extract the generalized Baldin sum rule.

To calculate the integral, $F_1$ needs to be extracted from the measured cross sections in both the resonance and DIS regions. This requires accurate knowledge of the separated structure function $R$ in both regions, where $R=\sigma_L/\sigma_T$ is the ratio of longitudinal to transverse cross sections. In contrast to the high quality $R$ measurements available in the DIS regime, there were very few $R$ measurements in the resonance region prior to Jefferson Lab experiment E94-110 \cite{liang}. Therefore, no precise inclusive $F_1$ data in the resonance region was available either to constrain the resonance analyses (such as the MAID analysis \cite{mart}, etc.) or to accurately compute the low $Q^2$ generalized Baldin sum rule.      
           
E94-110 measured inclusive scattering of unpolarized electrons from a hydrogen target in Hall C at Jefferson Lab (JLab) \cite{liang}. The data was accumulated in the nucleon resonance region \cite{liang,liangphd}, as well as the elastic region with absolute normalizations to better than 2\% \cite{elasticR}. The kinematic settings of this experiment were chosen to extract the nucleon structure function $R$ using the Rosenbluth technique at $M\leq W \leq 2$~GeV. The $Q^2$ range covered by our data set was between 0.3 and 5 GeV$^2$. A complete description of the data analysis and systematic uncertainty estimations may be found in reference \cite{liang,liangphd,elasticR}. The longitudinal-transverse separations allowed the nucleon structure functions $F_2$, $F_1$ (purely transverse), and $F_L$ (purely longitudinal) to be extracted independently. Formally,
\begin{equation}
  \label{eq:R}
R={{F_2}\over{2xF_1}}\left(1+{{4M^2x^2}\over {Q^2}}\right)-1={{F_{L}}\over {2xF_{1}}}.  
\end{equation}

\begin{figure}
\includegraphics[angle=0,width=3in,height=3in]{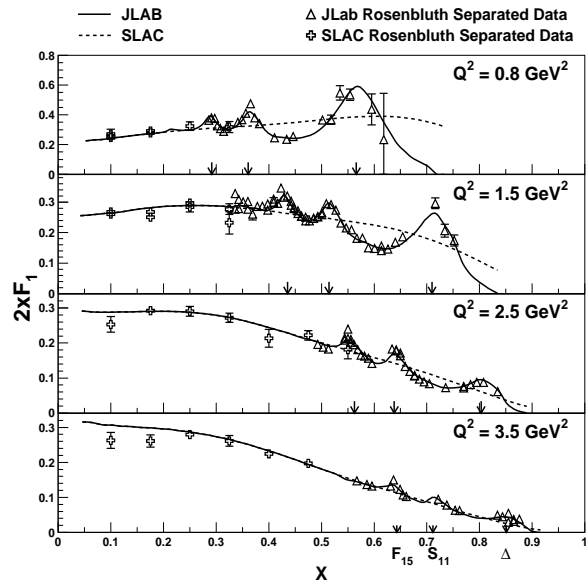}
\caption{\label{fig:baldinprl1}$2xF_1$ is plotted as a function of Bjorken $x$, at four different $Q^2$ values. The three arrows indicate where the three primary resonance enhancements are located.}
\end{figure}

A sample of the extracted $2xF_1$ data ($2xF_1 \sim \sigma_T$) in the nucleon resonance and DIS regions is shown in Fig.\ \ref{fig:baldinprl1}, as a function of Bjorken $x$ at four different $Q^2$ values. The open triangles in the figure represent the data extracted from the E94-110 Rosenbluth separations, and the open crosses represent the data extracted from SLAC Rosenbluth data \cite{whitlow}. The solid curve was calculated using the $R$ and $F_2$ parameterization of E94-110 \cite{liangphd} at $W<2$~GeV, and using the parameterization of $R$ and $F_2$ from SLAC DIS experiments \cite{whitlow,abe} at $W>2$~GeV. Note that the solid curve well reproduces the data in both the resonance and DIS regions. The dashed curve in the figure was calculated using only the parameterization of $R$ and $F_2$ from SLAC DIS experiments \cite{whitlow,abe}. A comparison between the JLab E94-110 and SLAC data sets was done at $3.5 <W^2 <4$~GeV$^2$, and the two data sets were found to be consistent within 1\% in this overlap region \cite{liangphd}. 

We calculate the second moment of $F_1$ by integrating the area below the solid curve over the Bjorken $x$ range $0< x <x_0$. The area corresponding to $W<2$ represents the resonance contribution to the moment, while the area corresponding to $W>2$ is the DIS contribution. Using this approach, we separated the generalized Baldin sum rule into two pieces, contributions from the resonance and DIS regimes,
\begin{eqnarray}
  \label{eq:exbaldin1}
 {\alpha(Q^2)+\beta(Q^2)}&=&{{{e^2M}\over {\pi{Q^4}}}\int_{x_{res}}^{x_0}{2xF_1(x,Q^2)}\, dx} \nonumber\\ 
& &+{{{e^2M}\over {\pi{Q^4}}}\int_{0}^{x_{res}}{2xF_1(x,Q^2)}\, dx},   
\end{eqnarray}
where $x_{res}$ corresponds to $W=2$~GeV. The uncertainty of this extracted generalized sum rule is less than 3\%, dominated by the normalization systematic uncertainties of the measured cross sections ($\sim$2\%) \cite{liangphd,whitlow}, as well as fitting uncertainties ($\sim$2\%) \cite{liangphd,whitlow,abe}.

Table\ \ref{tab:table1} lists the generalized sum rule values at eleven different $Q^2$ values, as well as the resonance and DIS contributions separately. Note that the generalized Baldin sum rule extracted from this experiment is not the same as the sum rule extracted from virtual Compton scattering \cite{laveissiere}. While these two quantities should converge at $Q^2=0$, the measured process of this experiment involves two virtual photons, as compared to one virtual photon in virtual Compton scattering. The latter $\alpha+\beta$ value is $1.51 \pm 0.25$~($10^{-4}$~fm$^3$) at $Q^2 =0.92$~GeV$^2$ \cite{laveissiere}, whereas this result is $0.4367 \pm 0.0131$~($10^{-4}$~fm$^3$) at $Q^2=1.0$~GeV$^2$. 

\begin{table}
\vspace{0.8cm}
  \begin{center} 
\begin{ruledtabular}
   \begin{tabular}{cccc}
$Q^2$~(GeV$^2$) & Sum Rule ($10^{-4}fm^3$)  & Resonance & DIS\\ \hline
0.3     & 3.0673 & 2.7137 & 0.3536 \\ 
0.4     & 2.2444 & 1.9513 & 0.2931 \\ 
0.6     & 1.2033 & 0.9894 & 0.2139 \\ 
0.8     & 0.6945 & 0.5306 & 0.1639 \\ 
1.0     & 0.4367 & 0.3053 & 0.1314 \\ 
1.5     & 0.1813 & 0.0979 & 0.0834 \\ 
2.0     & 0.0972 & 0.0399 & 0.0573 \\ 
2.5     & 0.0604 & 0.0188 & 0.0416 \\ 
3.0     & 0.0412 & 0.0099 & 0.0313 \\ 
3.5     & 0.0299 & 0.0055 & 0.0243 \\ 
4.0     & 0.0226 & 0.0032 & 0.0194 \\ 
    \end{tabular}
\end{ruledtabular}
     \caption{Table of the generalized Baldin sum rule values, as well as the resonance and DIS contributions to the sum rule, for the $Q^2$ values given. The uncertainty of the calculated sum rule is estimated to be less than 3\% (see text).}
   \label{tab:table1}
  \end{center}
\end{table} 

\begin{figure}
\vspace{0.5cm}
\includegraphics[angle=0,width=3in,height=3in]{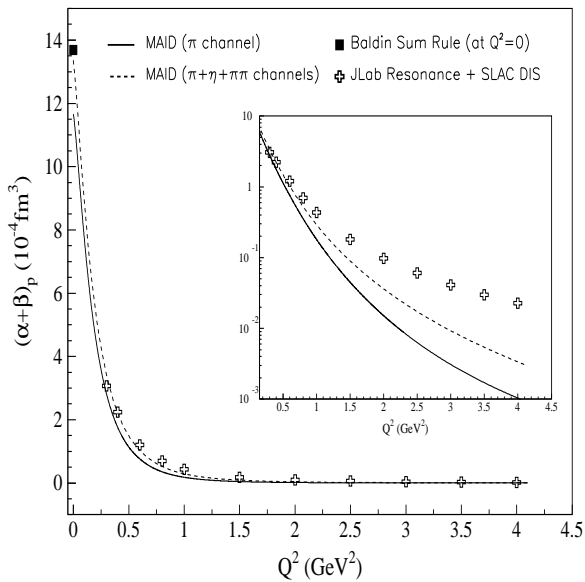}
\caption{\label{fig:baldinprl2}A comparison of MAID estimates with the extracted sum rule value at different $Q^2$.}
\end{figure}

The extracted extended sum rule value is plotted as a function of $Q^2$ in Fig.\ \ref{fig:baldinprl2}, along with the Baldin sum rule at $Q^2=0$. It clearly shows that, unlike the generalized GDH sum rule, the generalized Baldin integral evolves smoothly to the $Q^2=0$ point as has been predicted\ \cite{drechsel}. We also plot two MAID estimates, one for the $\pi$ channel only, and another for the $\pi+\eta+\pi\pi$ channels\ \cite{drechsel1,drechsel2}, as a comparison to our measurement. The latter agrees better with our data, clearly indicating that accurate modeling of this integral requires multiple resonance contributions. The discrepancy between the data and MAID estimate at $Q^2>1$~GeV$^2$ (see figure inset) is likely due to DIS and higher mass contributions of the extracted integral.      

\begin{figure}
\vspace{0.5cm}
\includegraphics[angle=0,width=3in,height=3.0in]{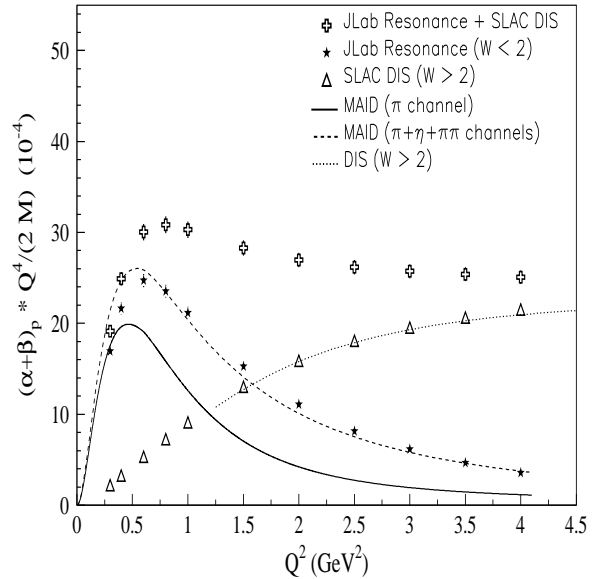}
\caption{\label{fig:baldinprl3}A comparison of MAID estimates with the generalized Baldin integral weighted by a factor of ${Q^4/ 2M}$, at different $Q^2$, as well as the resonance and DIS contributions.}
\end{figure}

To compare the $Q^2$ evolution of the resonance and DIS contributions of the generalized Baldin integral, we plot the sum rule value, resonance and DIS contributions in Fig.\ \ref{fig:baldinprl3}, each multiplied by a factor of $Q^4/2M$, as a function of $Q^2$, along with the two MAID estimates, and one DIS estimate calculated from the SLAC parameterization \cite{whitlow,abe}. The figure shows that the resonance contribution extracted from our data set agrees well with the three channel MAID estimate down to $Q^2=0.6$~GeV$^2$. 

The generalized sum rule is mainly saturated by the resonance contribution at $Q^2\leq 1$~GeV$^2$, while the DIS part dominates at $Q^2\geq 2$~GeV$^2$. At $1< Q^2 <2$~GeV$^2$, a transition from partonic incoherent processes to resonance dominated coherent processes occurs. Also, the value of $Q^4(\alpha+\beta)/2M$, which is proportional to the second moment of structure function $F_1$, is nearly flat at $Q^2>2.5$~GeV$^2$. This behavior is predicted by the perturbative description of DIS processes at large $Q^2$ \cite{rujula, rujula1}. 

In summary, we have utilized new inclusive electron-proton scattering cross section results in the resonance region, allowing for extraction of the structure functions $R$, $F_2$, $F_L$ and $F_1$. The $F_1$ data was used to calculate the generalized Baldin sum rule over the $Q^2$ range from 0.3 to 4.0 GeV$^2$. A transition from partonic incoherent processes to resonance dominated coherent processes is observed at $Q^2$ between 1 and 2 GeV$^2$. Resonance models of this integral need to include higher mass contribution. The generalized Baldin sum rule is found to evolve smoothly with $Q^2$ to the real photon point. 

The authors would like to thank Charles Hyde-Wright for many useful discussions. This work was supported in part by research grants 0244999, 0099540, and 9633750 from the National Science Foundation. The Southeastern Universities Research Association operates the Thomas Jefferson National Accelerator Facility under U.S. Department of Energy contract DEAC05-84ER40150.


\begin{thebibliography}{99}
\bibitem{baldin} A.M. Baldin, Nucl. Phys. {\bf 18}, 310 (1960)
\bibitem{lapidus} L.I. Lapidus, Sov. Phys. JETP {\bf 16}, 964 (1963)
\bibitem{babusci} D. Babusci {\it et al.}, Phys. Rev. C {\bf 57}, 291 (1998)
\bibitem{gerasimov} S. Gerasimov, Sov. J. Nucl. Phys. {\bf 2}, 430 (1966)
\bibitem{drell} S.D. Drell and A.C. Hearn, Phys. Rev. Lett. {\bf 16}, 908 (1966)
\bibitem{anselmino} M. Anselmino {\it et al.}, Sov. J. Nucl. Phys. {\bf 49}, 136 (1989)
\bibitem{kronig} R.de L. Kronig, J. Opt. Soc. Amer. Rev. Sci Instrum. {\bf 12}, 547 (1926); \\H.A.Kramers, Atti Congr. Int. Fis. Como. {\bf 2}, 545 (1927)
\bibitem{low} F.E. Low, Phys. Rev. {\bf 96}, 1428 (1954) 
\bibitem{drechsel} D. Drechsel, B. Pasquini, and M. Vanderhaeghen, Phys. Rep. {\bf 378}, 99 (2003)
\bibitem{brand} J.F.J. van den Brand {\it et al.}, Phys. Rev. D {\bf 52}, 4868 (1995)
\bibitem{roche} J. Roche {\it et al.}, Phys. Rev. Lett. {\bf 85}, 708 (2000)
\bibitem{jaminion} S. Jaminion {\it et al.}, hep-ph/0312293, submited to Phys. Rev. Lett. (2004)
\bibitem{laveissiere} G. Laveissiere {\it et al.}, hep-ph/0312294, submited to Phys. Rev. Lett. (2004)
\bibitem{cornwall} J.M. Cornwall and R.E. Norton, Phys. Rev. {\bf 177}, 2584 (1969)
\bibitem{hand} L.N. Hand, Phys. Rev. {\bf 129}, 1834 (1963)
\bibitem{callan} C.G. Callan {\it et al.}, Phys. Rev. Lett. {\bf 21}, 311 (1968)
\bibitem{rujula} A. De Rujula, H. Georgi, and H.D. Politzer, Ann. Phys. {\bf 103}, 315 (1977)
\bibitem{rujula1} A. De Rujula, H. Georgi, and H.D. Politzer, Phys. Lett. {\bf B64}, 428 (1977)
\bibitem{liang} Y. Liang {\it et al.}, nucl-ex/0410027, submitted to Phys. Rev. Lett. (2004)
\bibitem{mart} T. Mart {\it et al.}, http://www.kph.uni-mainz.de/MAID/
\bibitem{liangphd} Y. Liang, Ph.D. thesis, The American University (2003).
\bibitem{elasticR} M.E. Christy  {\it et al.}, Phys. Rev. C {\bf 70}, 015206 (2004)
\bibitem{whitlow} L.W. Whitlow {\it et al.}, Phys. Lett. {\bf B250}, 193 (1990)
\bibitem{abe} K. Abe {\it et al.}, Phys. Lett. {\bf B452}, 194 (1999)
\bibitem{drechsel1} D. Drechsel {\it et al.}, Nucl. Phys. {\bf A645}, 145 (1999)
\bibitem{drechsel2} D. Drechsel {\it et al.}, Phys. Rev. D {\bf 63}, 114010 (2001)
\end{thebibliography}
\end{document}